\documentclass{emulateapj}

\usepackage{graphicx}
\usepackage[toc,page]{appendix}
\usepackage{amssymb}
\usepackage{amsmath}
\usepackage{array}

\newcommand{\eg}{e.g., }

\newcommand{\etal}{et al.}

\newcommand{\code}[1]{\texttt{#1}} 
\newcommand{\degree}{\ensuremath{^\circ}}

\begin{document}

\title{Numerical Simulations of Saturn's B Ring:\\
	Granular Friction as a Mediator between Self-Gravity Wakes and Viscous Overstability}
\author{Ronald-Louis Ballouz$^1$, Derek C. Richardson$^1$, Ryuji Morishima$^{2,3}$}
\affil{$^1$Department of Astronomy, University of Maryland, College Park, MD, 20742-2421, USA\\
       $^2$University of California, Los Angeles, Institute of Geophysics and Planetary Physics, Los Angeles, CA, USA\\
       $^3$Jet Propulsion Laboratory/California Institute of Technology, Pasadena, CA, USA}


\begin{abstract}
We study the B ring's complex optical depth structure. The source of this structure may be the complex dynamics of the Keplerian shear and the self-gravity of the ring particles. The outcome of these dynamic effects depends sensitively on the collisional and physical properties of the particles. Two mechanisms can emerge that dominate the macroscopic physical structure of the ring: self-gravity wakes and viscous overstability. Here we study the interplay between these two mechanisms by using our recently developed particle collision method that allows us to better model the inter-particle contact physics. We find that for a constant ring surface density and particle internal density, particles with rough surfaces tend to produce axisymmetric ring features associated with the viscous overstability, while particles with smoother surfaces produce self-gravity wakes.
\end{abstract}


\section{Introduction}
Observations have shown that Saturn's rings exhibit rich and varied structure. Observations from the Cassini spacecraft show that this structure exists on many different scales, from 100's of meters to 100's of kilometers (Porco et al.\ 2005, Colwell et al.\ 2007, 2009). The B ring, in particular, exhibits a varied and complex structure as uncovered by photometric optical depth measurements (Colwell et al.\ 2009). This structure likely comes about from the combined effects of ring particle collisions, ring self-gravity, and tidal forces of Saturn. For the origin of structure of sub-km scales, two mechanisms have been proposed.

The first of these mechanisms is the formation of gravitational wakes in the dense rings. Self-gravity causes the ring particles to clump, forming transient structures that are quickly disrupted by the background shear. These wakes have typical wavelengths of a few 100's of meters, and they are chiefly characterized by their pitch angle of 10$\degree$--30$\degree$, the angle that the opaque clumps form with respect to the azimuthal (orbital) direction of motion. While they have never been imaged directly, their presence have been inferred from Cassini Ultraviolet Imaging Spectrograph (UVIS) and Visual and Infrared Mapping Spectrometer (VIMS) optical depth measurements of the A and B rings (Colwell et al.\ 2006, Hedman et al.\ 2007).

The self-gravitational interactions of the particles create a dense packing of particles that are able to dissipate energy through inelastic collisions, leading to an increase in the angular momentum transport in the ring. For a sufficient number of dissipative collisions, the viscosity of the ring may evolve to the point that a viscous overstability develops (Salo et al.\ 2001). The second mechanism for structure formation develops when the ring viscosity steeply increases with the surface density due to frequent dissipative collisions. This process is driven by particles in over-dense regions flowing towards under-dense regions, as in a viscously stable ring.  However, this smoothing process overshoots as particles moving radially away from the over-dense structure collide with particles moving radially outwards from an adjacent structure. This allows new over-dense regions to develop. These pulsations in ring density generate a wave pattern characterized by transient axisymmetric over-dense regions (regions of high vertical coherence; see Hedman et al.\ 2014). Similar to gravity wakes, these opaque clumps have sub-km wavelength, typically $\sim$ 100 m (Salo et al.\ 2001, Daisaka et al.\ 2001, Rein \& Latter 2013). Cassini infrared occulation measurements of the B ring (Colwell et al.\ 2007) show that sub-km structure exists and appears to be structurally stable and axisymmetric at scales of up to 3,000 km, many orders of magnitude larger than the radial overstability wavelength, suggesting that such overstable structures do indeed exist in the B ring.

The development of the viscous overstability depends sensitively on the collisional dynamics of the ring particles. Initially, a viscous instability was hypothesized to explain the structure in the B ring. In this scenario, particle collisions were assumed to be highly elastic, which causes the dynamic shear viscosity of the ring to decrease with density, and the particle flux to be directed towards regions of high density. This also results in the formation of banded structure; however, these structures persist indefinitely. Laboratory experiments by Bridges et al.\ (1984) showed that collisions between frictionless and spinless icy spheres at impact speeds typical of Saturn's dense rings (a few mm/s) are actually highly inelastic. Subsequent laboratory and numerical studies showed the first hints that the viscous overstability could be possible in planetary rings (Borderies et al.\ 1985, Salo 2001). These laboratory experiments (Bridges et al.\ 1984, Borderies et al.\ 1985) revealed that the restitution coefficient of the ice spheres was a function of the impact speed. This functional dependence of the restitution coefficient with the impact speed predicts an energy equilibrium that allows for the flow of ring material to be viscously unstable or overstable, depending on the ring particles' exact properties.\ 

The study of the self-gravity and viscous overstability mechanisms has been advanced through numerical simulations using $N$-body integrators that simulate the collisional and gravitational evolution of some 10's to 100's of thousands of particles in a co-moving patch (e.g., Salo 1992, Yasui et al.\ 2012, Rein \& Latter 2013). These studies employ the functional dependency of the dissipative parameters on the impact speeds found in Bridges et al.\ (1984) to model the collisions of hard-sphere particles. In the hard-sphere method, particle collisions occur as pair-wise interactions that are resolved instantaneously. These simulations led to predictions that the viscous overstability may only develop when the internal density of ring particles is low (200--300 kg/m$^3$), implying the particles are highly porous. At higher particle densities, the viscosity of the ring is insufficient to keep it stable against self-gravitational perturbations, and gravity wakes develop. Studies have shown that both viscous overstability and self-gravity wakes may co-exist for certain ring physical and collisional properties (Salo et al.\ 2001). However, the exact nature of the relationship between these two mechanisms is not well known. While nonaxisymmetric wakes suppress the growth of axisymmetric overstable oscillations, the ring self-gravity is still a necessary component to the emergence of overstability features. Self-gravity contributes a vertical component to particle motion. Combined with dissipative collisions, this initial vertical motion focuses particles to the disk mid-plane, which, in turn, leads to more collisions. There is currently no analytical theory that is able to model the formation of overstability in a ring with self-gravity wakes.

In this study, we perform direct numerical simulations of a patch of particles in Saturn's dense B ring. Our $N$-body code allows us to study the collisional and gravitational dynamics in high optical depth regions in Saturn's rings. Using a recently implemented collisional method that allows the self-consistent modeling of multi-contact and frictional forces (Schwartz et al.\ 2012), we set out to better understand the interplay between the two mechanisms that create the structure of Saturn's B ring, with the goal of generating a library of synthetic observations of the optical depth structure mapped to particle physical and collisional properties. In Section 2 we describe our numerical method and how we compute viscosity in the simulations.  In Section 3 we show the results of our sweep over density and friction parameters that appear to determine equilibrium optical depth structure in Saturn's B ring. In Section 4 we present our conclusions and suggestions for further study.

\section{Methodology}

\subsection{Numerical Method}
We use \code{pkdgrav}, a parallel $N$-body gravity tree code (Stadel 2001) adapted for particle collisions (Richardson \etal\ 2000; 2009; 2011).  Originally collisions in \code{pkdgrav} were treated as idealized single-point-of-contact impacts between rigid spheres.  A soft-sphere option was added recently (Schwartz \etal\ 2012); with this option, particle contacts can last many time steps, with reaction forces dependent on the degree of overlap (a proxy for surface deformation) and contact history. This allows us to model multi-contact and frictional forces. The code uses a second order leapfrog integrator, with accelerations due to gravity and contact forces recomputed each step.

The soft-sphere implementation in \code{pkdgrav} uses a spring/dash-pot to model the collisional forces between particles. In this model, a spherical particle overlapping with a neighboring particle feels a reaction force in the normal and tangential directions determined by spring constants ($k_n$, $k_t$), with optional damping and effects that impose static, and/or rolling friction. The damping parameters ($C_n$, $C_t$) are related to the conventional normal and tangential coefficients of restitution used in hard-sphere implementations, $\varepsilon_n$ and $\varepsilon_t$.  An option to have an impact-speed dependent $\varepsilon_n$ is also available. This option was used to benchmark our simulation results against previous studies (see Section 2.3). The static and rolling friction components are parameterized by dimensionless coefficients $\mu_s$ and $\mu_r$, respectively.

The coefficient of static friction determines the maximum amount of tangential force that can be supported by the contact point of two particles. This tangential force scales linearly with the normal force at the contact point. In the event that the tangential force exceeds the maximum value given by the static friction coefficient, the particles slip past one another, and the tangential force is then determined by the tangential spring constant and the tangential damping coefficient.

The coefficient of rolling friction is used to determine the resulting torque due to two particles rolling against one another. Two particles are said to be rolling with respect to each other if their relative velocities are zero, but have some relative rotational motion. The torque induced by the rotation scales linearly with coefficient of rolling friction. In practical terms, a non-zero coefficient of rolling friction mimics some asphericity in the grains, increasing the bulk resistance to shear. Plausible values for these various parameters are obtained through comparison with laboratory experiments (Section 2.2).

Our numerical approach has been validated through comparison with laboratory experiments; \eg Schwartz \etal\ (2012) demonstrated that \code{pkdgrav} correctly reproduces experiments of granular flow through cylindrical hoppers, specifically the flow rate as a function of aperture size, Schwartz \etal\ (2013) demonstrated successful simulation of laboratory impact experiments into sintered glass beads using a cohesion model coupled with the soft-sphere code in \code{pkdgrav}, and Schwartz et al.\ (2014) applied the code to low-speed impacts into regolith in order to test asteroid sampling mechanism design.

Since soft-sphere discrete element method models treat particle collisions as reactions of springs due to particle overlaps, the magnitude of the normal and tangential restoring forces are determined by the spring constants $k_n$ and $k_t$ (although most implementations, including ours, conventionally set $k_t$ $\sim$ $\frac{2}{7}$ $k_n$ to keep the damped harmonic frequencies of the normal and tangential springs in phase---see Schwartz et al.\ (2012). An appropriate value of $k_n$ is given by
\begin{equation} k_n \sim m \left(\frac{v_{max}}{x_{max}}\right)^{2},\end{equation}
\noindent where $m$ is the mass of a particle, $v_{max}$ is the maximum expected relative particle speed in the simulation, and $x_{max}$ is the maximum allowed particle overlap, which we set to be $\sim 1\%$ of the particle radius.  In the case of a particle size distribution, the maximum particle mass is used for $m$ and the minimum particle radius is used to determine $x_{max}$, thereby giving the most conservative estimate for $k_n$, although for the results reported here all particles in a given simulation had identical properties. The time step is set so that a typical collision (i.e., a full oscillation of the normal spring) takes about 30 steps.

We simulate the collisions of ring particles in Saturn's B ring, a distance of 100,000 km away from Saturn. The typical collision speeds of particles due to Keplerian shear is expected to be of the order of a few times $\Omega r_{p}$, where $\Omega$ is the Keplerian orbital angular frequency for an object orbiting Saturn at 100,000 km ($\sim$ $2\times10^{-4}$ $\mathrm{s}^{-1}$), and $r_{p}$ is the radius of a ring particle. Thus, for our ring simulations, particles typically collide at relative speeds of $\sim 0.5-1$ $\mathrm{mm}$ $\mathrm{s}^{-1}$. Therefore, we set $k_n$ $\sim 4.71$ $\mathrm{kg}$ $\mathrm{s}^{-2}$ and the time step $\sim$ $1.5$ $\mathrm{s}$. The collisional model described here (Schwartz et al.\ 2012) is not the first application of a soft-sphere model to planetary ring simulations. Salo et al.\ (2001) presented the first simulations to show the emergence of overstable oscillations in a ring with self-gravitating ring particles, using a spring force method described in Salo (1995). However, the inclusion of a tangential spring constant and rolling friction in the force model used in this work is novel. 

\subsection{Periodic Boundary Conditions \& Patch Properties}
We perform local simulations of ring particles by restricting the computational volume to a small region (a ``patch'') of the entire ring. Particles are modeled in a co-moving patch frame. The linearized equations of motion, called the Hill's equations, are (Hill 1878; also see Wisdom \& Tremaine 1988):

\begin{eqnarray*}
\ddot{x} & = & F_x + 3 \Omega^2 x + 2 \Omega \dot{y},\\
\ddot{y} & = & F_y - 2 \Omega \dot{x},\\
\ddot{z} & = & F_z - \Omega^2 z,
\end{eqnarray*}

\noindent where $\mathbf{F}=(F_x,F_y,F_z) $ is the acceleration due to particle self-gravity, $x$, $y$, and $z$ are the coordinates of the particle in the local coordinate system (whose origin is located at the center of the patch), and the derivatives are with respect to time. We use periodic boundaries in the plane of the patch. Periodic boundary conditions (PBCs) are used to model a patch of the ring, as modeling the entire ring system would be computationally infeasible. 

\begin{figure}
\begin{center}
\includegraphics[width=0.5\textwidth]{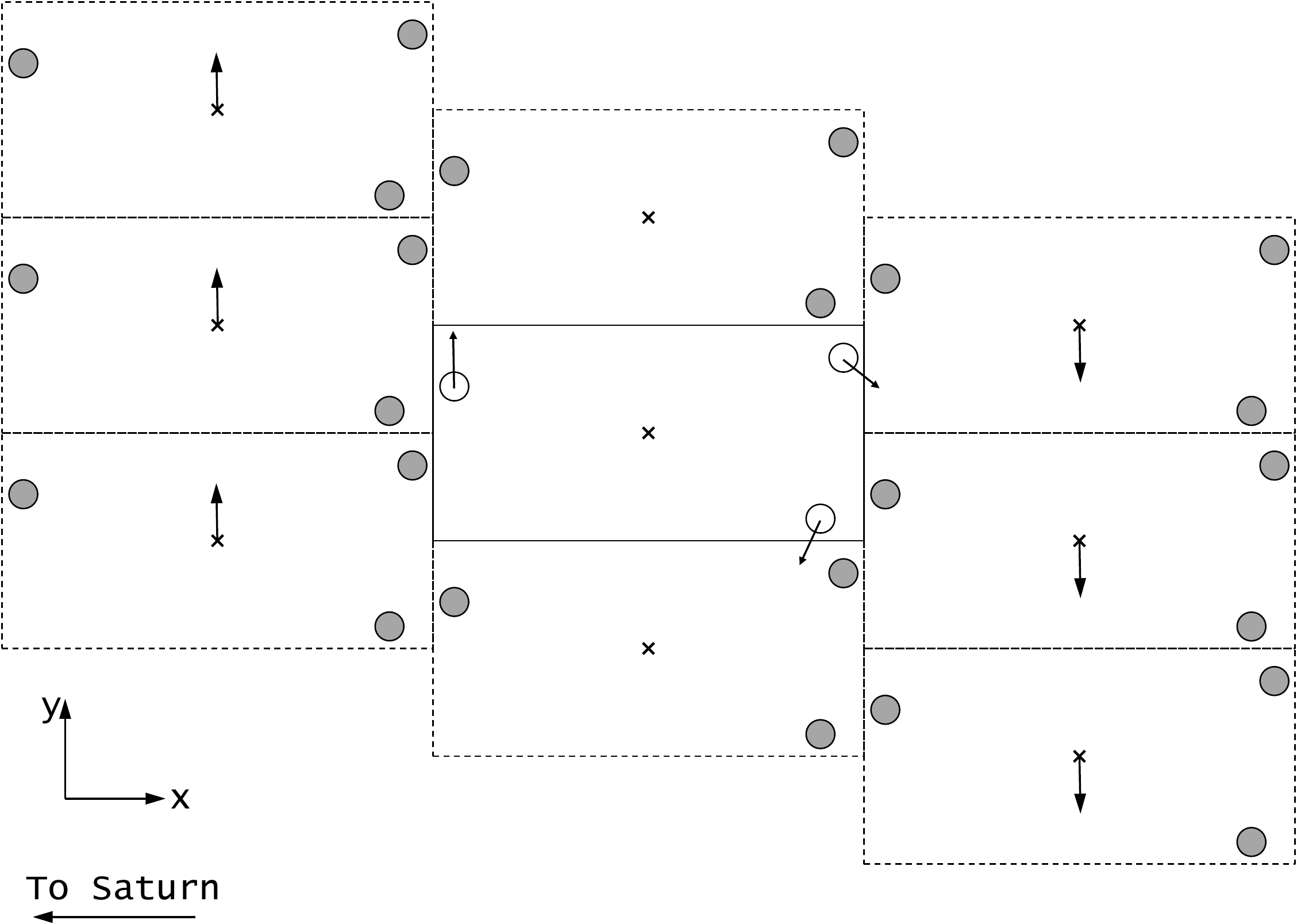}
\caption{\footnotesize{
A schematic diagram of a local sliding-patch model with shearing periodic
boundary conditions. The three white particles in the center solid box are the simulated particles; grey particles are in replicated
patches (dashed boxes) that provide boundary conditions. The $x$ coordinate is the radial direction, with Saturn located in the negative $x$ direction; $y$ is the azimuthal direction, and the entire patch orbits Saturn in the positive $y$ direction. $z$ points out of the page, forming a right-handed coordinate system. The simulation is carried out in the orbital frame of the center of the patch, so particles to the left shear upward (positive $y$ direction); on the right, they shear downward (negative $y$ direction). The replicated patches similarly shear past the center box in Keplerian fashion; each bold X marks the center of each patch, with the bulk relative motion of each patch indicated by black arrows.}}\label{RingPBC}
\end{center}
\end{figure}

PBCs are implemented by replicating a patch of particles in the radial ($x$) and azimuthal ($y$) directions. Each replicated patch contains ``ghost'' particles that match the relative positions of the original particles. Figure~\ref{RingPBC} illustrates the PBC setup. Particles in the central patch feel the gravitational effects of ghost particles, and can collide with ghost particles at the boundaries. The azimuthal and radial extents of the patch are small compared to the orbital distance to the planet, but large compared to the radial mean free path of the particles inside the patch. Particles exiting one side of the azimuthal or radial boundary reappear on the other side. When a particle crosses the radial boundary, its azimuthal speed is modified to account for the Keplerian shear. Therefore, depending on whether it crosses over from a smaller radial position to a larger one or vice versa, its azimuthal speed will be adjusted by $\pm \frac{3}{2} \Omega L_{x}$, where $L_{x}$ is the radial width of the patch.

We simulate a patch of the B ring that has a dynamic optical depth of 1.4 and a surface density of 840 kg m$^{-2}$. The dynamical optical depth $\tau_{\mathrm{dyn}}$ of a ring is defined as
\begin{equation} \tau_{\mathrm{dyn}} = \int \pi s^2 n(s) \mathrm{d}s,\end{equation}
where $s$ is the particle radius and $n(s)$ is the surface number density of ring particles whose sizes are between $s$ and $s+ds$. We choose a value of the optical depth of 1.4 in our simulations as it is typical of the B ring, and it corresponds to a number of particles that can be feasibly simulated in a timely fashion using our code. The value of the surface density is well within the range found through analysis of density waves (400--1400 kg m$^{-2}$ (Hedman \& Nicholson 2016). In order to ensure that we allow sufficient space for the formation of large-scale structures, we simulate a patch that is at least $20$ $\lambda_{cr}$ in the radial direction and $4$ $\lambda_{cr}$ in the azimuthal direction, where  $\lambda_{cr}$ is the Toomre wavelength, defined as:
\begin{equation} \lambda_{cr} = \frac{4\pi^2 G\sigma}{\Omega^2},\end{equation}
where $G$ is the gravitational constant and $\sigma$ is the surface density. Previous simulations (Salo 1992, Daisaka \& Ida 1999) have shown that the typical wavelength of wakes is close to $\lambda_{cr}$.

For each simulation, we choose a single combination of a particle density and particle radius (monodisperse distribution) that corresponds to a constant optical depth of 1.4 and a patch surface density of 840 kg m$^{-2}$. This is done in order to ensure $\lambda_{cr}$ is kept constant across all simulations, allowing us to draw more accurate comparisons of simulation outcomes across different particle density cases. Therefore, a simulation can contain particles with densities of 0.45, 0.60, 0.75, or 0.90 g/cm$^3$ that have radii of 1.0, 0.75, 0.6, and 0.5 m, respectively. The number of particles in our simulations ranges from  $\sim$120,000 to  $\sim$500,000 (see Table~\ref{t:summary_results}).

\subsection{Diagnostic Tools}

We developed diagnostic tools to study the bulk behavior of the particles in the patch and to compare our results to observations. The first tool calculates the viscosity of the patch, in order to measure the different contributions to the angular momentum transport in the ring. This is important for characterizing the influence of the individual ring properties to the larger-scale dynamics, and it allows us to directly compare our results to previous theoretical treatments of ring dynamics. The second tool uses a Monte Carlo ray-tracing algorithm to measure the photometric optical depth of our simulations. This allows us to directly compare the outcome of our simulations to Cassini results by producing synthetic stellar occultation data.

\subsubsection{Viscosity Calculation}

In a planetary ring system, the viscosity of the disk described how efficiently the system is able to transport angular momentum from interior parts to its exterior. In its simplest sense, viscosity is a measure of a fluid system's resistance to stresses, either shearing or tensile. For a ring system such as Saturn's, there are are three components that contribute to the viscosity in a disk: translational viscosity, collisional viscosity, and gravitational viscosity (Daisaka et al.\ 2001).The translational viscosity, $\nu_{\mathrm{trans}}$, is due to Keplerian shear and can be calculated by measuring the radial velocity dispersion in the ring patch:
\begin{equation}\nu_{\mathrm{trans}} = \frac{2}{3 \Omega \Sigma_{i} m_{i}} \Sigma_i m_i \dot{x}_i \dot{y}_{r,i}, \end{equation} 
\noindent where the sums are over all particles, $m_i$ is the particle mass, $\dot{x}_i$ is the $x$-component of the velocity, and $\dot{y}_{r,i}=\dot{y}_i + (3/2)\Omega x_i$ is the $y$-component of the velocity relative to the mean shear speed at radial position $x_i$.

The second component of the viscosity is due to the transfer of angular momentum by collisions. Previous authors (Wisdom \& Tremaine 1988, Daisaka et al.\ 2001) used event-driven hard-sphere collision codes. Thus, they calculate the exchange in momentum in the azimuthal direction for each discrete collision. The collisional component of the viscosity is given by:
\begin{equation} \nu_{\mathrm{coll}} = \frac{2}{3 \Omega M \delta T} \Sigma \delta p_{y} (x_{>} - x_{<}) ,\end{equation} 
\noindent where $M$ is the total mass in the patch. For each pairwise collision that occurs during a time interval $\delta T$, $x_{>}$ is the radial distance of the particle with the exterior orbit, $x_{<}$ is the radial distance of a particle with the interior orbit, and $\delta p_{y}$ is the change in the azimuthal component of the momentum of the exterior particle. The summation is over all collisions that occur during the interval $\delta T$.

The third component of the viscosity is due to the self-gravitational interaction of ring particles. While this is an important contribution to the overall viscosity of a ring system, we do not measure this component in this study. Rather, we focus on how collisional interactions of the disk are affected by different surface properties of the particles. We refer the reader to Daisaka et al.\ (2001) for a full discussion on gravitational viscosity. 

Since our soft-sphere code treats collisions as finite duration with possible multiple simultaneous contacts and frictional forces, we must revise the formulation for the collisional shear. Following the treatment in Wisdom \& Tremaine (1988), the exchange of momentum between two particles can be obtained from collisional force balance. Therefore, in a soft-sphere collision implementation, the collisional viscosity can be calculated as:
\begin{equation}\nu_{\mathrm{coll}} = \frac{2}{3 \Omega M}  \Sigma  F_{y >} (x_{>} - x_{<}),\end{equation}
\noindent where $F_{y>}$ is the collisional force in the azimuthal direction on a particle due to its colliding partner(s). The summation is over all colliding particle pairs at each timestep. Formulated in this way,  we are able to take into account multi-contact and frictional forces in the viscosity calculation. 

Another method that involves measuring the change in orbital elements due to collisions has been devised to measure the total viscosity (sum of translational, collisional, and gravitational) in the ring patch (Tanaka et al.\ 2003). Yasui et al.\ (2012) recently used this method to calculate the viscosity of spinning self-gravitating particles in planetary rings. They were also able to take the first steps in measuring the effect of friction on the viscosity by including the effects of a tangential restitution coefficient. Yasui et al.\ (2012) demonstrated the validity of their calculation by performing simple tests to measure the viscosity in a small patch and compare their results to those found by Wisdom \& Tremaine (1988).

\begin{figure}
\begin{center}
\includegraphics[width=0.5\textwidth]{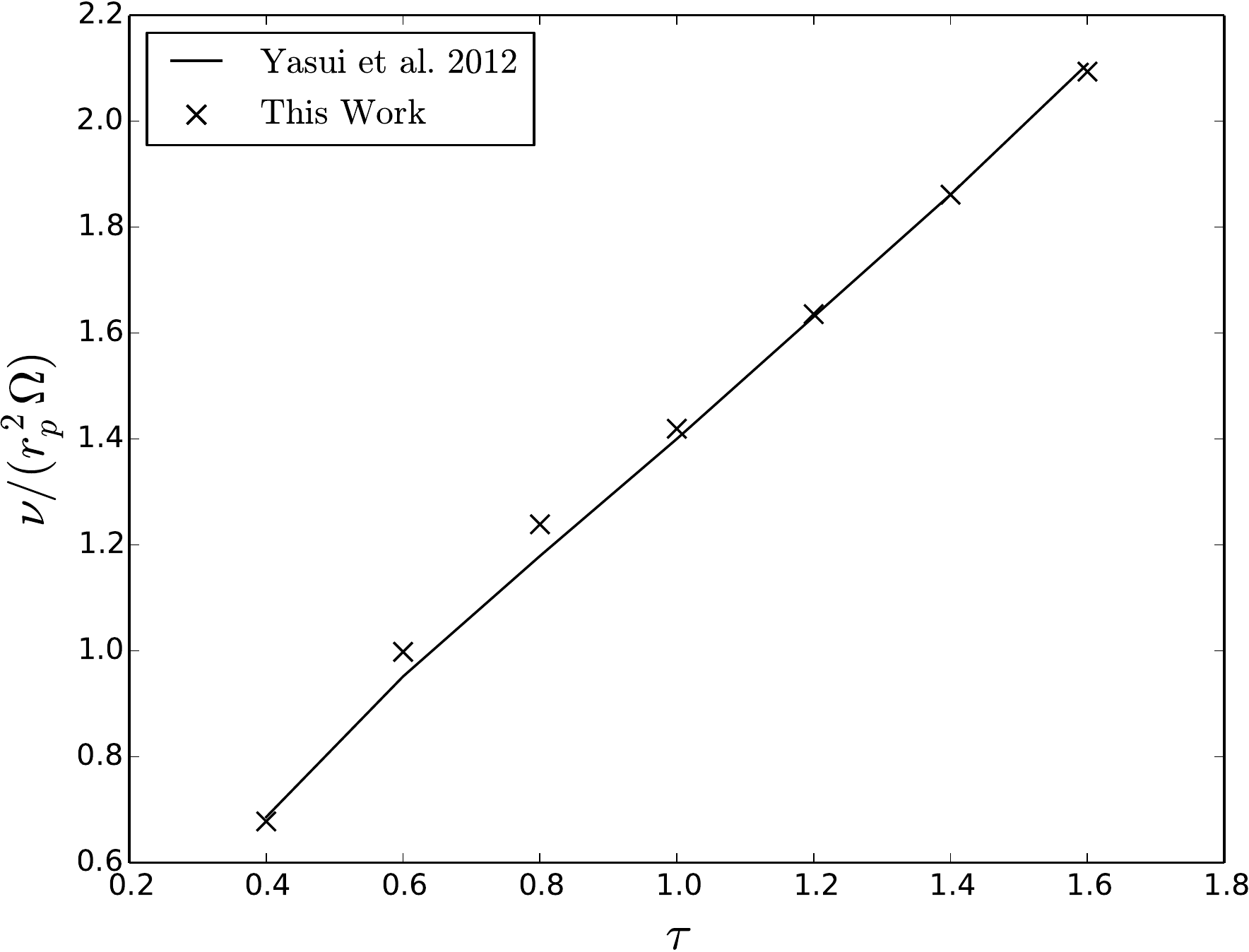}
\caption{\footnotesize{Total viscosity, $\nu$, normalized by $\Omega r_p^2$ as a function of the dynamical optical depth of the ring. Our total viscosity calculations (translational and collisional) match those found by Yasui et al.\ (2012) (see their Fig.\ 1b). The simulations are for a velocity-dependent normal restitution coefficient with $\epsilon_t$ = $0.5$.}}\label{yasui_comp}
\end{center}
\end{figure}

In order to validate our re-formulation of the Wisdom \& Tremaine (1988) prescriptions for calculating the collisional viscosity, and to ensure that our code is capable of replicating results found in recent numerical simulations of ring systems, we performed a simple simulation test similar to that done by Yasui et al.\ (2012). Using a square 125 m $\times$ 125 m ring patch made up of 1-m radius particles, we ran ring simulations with a speed-dependent $\epsilon_n$ derived from the laboratory impact experiments of Bridges et al.\ (1984), given by:
\begin{equation}\epsilon_n (v) = \mathrm{min} \{0.32(v/v_c)^{-0.234} , 1 \} ,\end{equation}
\noindent where $v$ is the relative speed between colliding particles at the moment of first overlap, and $v_c$ is a normalization equal to $1$ cm s$^{-1}$. We performed these simulations for ring patches with optical depths ranging from 0.4 to 1.6, and for a value of $\epsilon_t$ of 0.5. These runs had the self-gravity between particles turned off since we are only interested in the equilibrium viscosity achieved through ring-particle collisions. The simulations were run until the particles reached collisional equilibrium (a few 10's of orbits), determined by when the patch reached a steady radial velocity dispersion. We find excellent agreement between the viscosities we measure and those found in Yasui et al.\ (2012) (see Fig.~\ref{yasui_comp}).

\subsubsection{Photometric Optical Depth Calculation}
The photometric normal optical depth $\tau_{\mathrm{phot}}$ is defined as (e.g., Colwell et al.\ 2007)
\begin{equation}\label{eq_tau_phot} 
\tau_{\mathrm{phot}} = - \sin |B| \log \left( \frac{I}{I_0} \right),
\end{equation}
where $B$ is the elevation angle of the observer, $I$ is the measured intensity, and $I_0$ is the unocculted intensity of the light source.

If particles are randomly distributed in the plane and the mutual separation is large enough, $\tau_{\mathrm{phot}}$ becomes independent of observational geometry and coincides with $\tau_{\mathrm{dyn}}$ (the limit of classic radiative transfer). If the mutual separation is comparable to the particle radius, $\tau_{\mathrm{phot}}$ becomes larger than $\tau_{\mathrm{dyn}}$. This effect is called the volume filling factor effect (e.g., Salo and  Karjalainen 2003). If the ring structure is not horizontally uniform due to wakes or waves, $\tau_{\mathrm{phot}}$ depends on the azimuthal angle of the observer. Usually, $\tau_{\mathrm{phot}}$ has a minimum value when the line of sight is aligned to the wake direction (e.g., Colwell et al.\ 2006; Hedman et al.\ 2007).

We developed a ray-tracing code to derive $\tau_{\mathrm{phot}}$ using outputs of $N$-body simulations. The basic schemes employed in the code are similar to those given in Salo and Karjalainen (2003). Briefly, the algorithm is as follows:
\begin{itemize}
\item The simulated region is partitioned into two-dimensional sub-cells, with particles allocated by sub-cell.
\item A ray is generated by randomly choosing $x$ and $y$ coordinates in the ring midplane, where $z$=0, and forming the unit vector $\textbf{n}$ directed towards the light source. In the Hill coordinate system, $\textbf{n}$ is defined by its components:
\begin{eqnarray*}
n_x & = & \cos \theta_s \sin B_s,\\
n_y & = & \sin \theta_s \sin B_s,\\
n_z & = & \cos B_s,
\end{eqnarray*} 
\noindent where $\theta_s$ and $B_s$ are the azimuthal and elevation angle of the light source. 
\item We first check whether the ray coming from the light source intersects any sub-cells, and then check whether the ray intersects any particles within an intersected sub-cell. If multiple sub-cells can be intersected by a single ray, then these sub-cells are checked in the order of which is intersected first by the ray's trajectory.
\item If the ray passes through the simulated region without intersecting any particles, it may still intersect with a ghost particle. Therefore, if an intersection is still possible, the ray is regenerated with a new position given by the periodic boundary conditions of the shearing system. Intersections are still possible if 
\begin{equation}
p_z < \mathrm{max}(z_i + r_i)
\end{equation}
or
\begin{equation}
p_z > \mathrm{min}(z_i - r_i)
\end{equation} 
\noindent where $p_z$ is the $z$ coordinate of the ray, $z_i$ is the $z$ coordinate of any particle in the simulation, and $r_i$ is the corresponding particle radius. Therefore, the algorithm compares the $z$ coordinate of the ray to the maximum and minimum vertical extent of the simulated particles.
\item We continue generating randomly positioned rays until we numerically converge on a value for the photometric optical depth. The simulated photometric optical depth is found by setting $I_0$, in Eq.~\ref{eq_tau_phot}, to the number of rays produced, and $I$ to the number of rays that pass through the ring without intersecting any particles. 
\end{itemize}

\section{Results}

We have performed simulations that explore the effects of particle density and friction on the collisional dynamics of the ring particles and the subsequent large-scale structure evolution. Each simulation was run for at least 100 orbits in order for the patch to reach collisional equilibrium. Since the surface properties of actual ring particles are not well constrained, we explored a wide range of friction properties. This allowed us to characterize the different large-scale structure that could form. Table 1 lists the simulations that were performed. Here we also summarize the outcome of each simulation by providing the final collisional and translational viscosities. Furthermore, we also list the Toomre $Q$ of the ring (Toomre 1964), where $Q$ is defined as:
\begin{equation}  
Q = \frac{c_r \Omega}{\pi G \sigma}, 
\end{equation}
\noindent where $c_r$ is the radial velocity dispersion and $\sigma$ is the surface density. $Q$ attains a time-averaged equilibrium value of the order of 2 in the case of strong wake structure. Previous results have shown the emergence of strong wake structure when the radial velocity dispersion due to collisions and gravitational encounters does not exceed that corresponding to $Q \sim 2$ (Salo 1995, Salo et al.\  2001, Ohtsuki \& Emory 2000).

\begin{table}
\caption{\footnotesize{Summary of ring patch parameters and simulations results.}}
\label{t:summary_results}
\parbox{.45\linewidth}{\centering
\begin{tabular}{ccccccc}
\hline
\hline
$\rho$ (g cm$^{-3}$) & $r_p$ (m) & $\mu_{\mathrm{s}}$ & $\mu_r$ & $\nu_{\mathrm{coll}}$ & $\nu_{\mathrm{trans}}$ & $Q$ \\
\hline
0.45 & 1.0 & 0.0 & 0.0 & 5.70 & 8.29 & 1.34 \\
0.45 & 1.0 & 0.0 & 0.2 & 5.64 & 8.23 & 1.31 \\
0.45 & 1.0 & 0.2 & 0.0 & 8.88 & 14.12 & 1.86 \\
0.45 & 1.0 & 0.2 & 0.2 & 9.91 & 16.41 & 2.09\\
0.45 & 1.0 & 0.4 & 0.0 & 11.24 & 16.32 & 2.07 \\
0.45 & 1.0 & 0.4 & 0.2 & 18.89 & 30.81 & 2.65 \\
0.45 & 1.0 & 0.6 & 0.0 & 13.98 & 20.98 & 2.30 \\
0.45 & 1.0 & 0.6 & 0.2 & 17.81 & 28.21 & 2.58 \\
0.45 & 1.0 & 0.8 & 0.0 & 17.12 & 21.06 & 2.63 \\
0.45 & 1.0 & 0.8 & 0.2 & 20.75 & 40.96 & 3.44 \\
0.45 & 1.0 & 1.0 & 0.0 & 20.97 & 20.08 & 3.04 \\
0.45 & 1.0 & 1.0 & 0.2 & 23.35 & 40.86 & 3.22 \\
\hline
0.60 & 0.75 & 0.0 & 0.0 & 6.81 & 14.41 & 2.34 \\
0.60 & 0.75 & 0.0 & 0.2 & 6.98 & 15.65 & 2.25 \\
0.60 & 0.75 & 0.2 & 0.0 & 10.97 & 22.43 &2.53 \\
0.60 & 0.75 & 0.2 & 0.2 & 10.71 & 20.04 & 2.71 \\
0.60 & 0.75 & 0.4 & 0.0 & 10.96 & 21.10 & 2.48 \\
0.60 & 0.75 & 0.4 & 0.2 & 13.34 & 24.81 & 2.96 \\
0.60 & 0.75 & 0.6 & 0.0 & 7.80 & 15.07 & 3.05 \\
0.60 & 0.75 & 0.6 & 0.2 & 16.67 & 34.48 & 3.09 \\
0.60 & 0.75 & 0.8 & 0.0 & 10.51 & 20.48 & 2.67 \\
0.60 & 0.75 & 0.8 & 0.2 & 14.57 & 25.74 & 3.13 \\
0.60 & 0.75 & 1.0 & 0.2 & 11.82 & 21.10 & 3.32 \\
\hline
0.75 & 0.60 & 0.0 & 0.0 & 7.45 & 22.19 & 2.60 \\
0.75 & 0.60 & 0.0 & 0.2 & 7.45 & 20.62 & 2.73 \\
0.75 & 0.60 & 0.2 & 0.0 & 12.01 & 26.76 &  2.95 \\
0.75 & 0.60 & 0.2 & 0.2 & 14.03 & 28.32 & 3.14 \\
0.75 & 0.60 & 0.4 & 0.0 & 13.43 & 30.97 & 2.92 \\
0.75 & 0.60 & 0.4 & 0.2 & 16.52 & 32.17 & 3.22 \\
0.75 & 0.60 & 0.6 & 0.0 & 13.47 & 27.77 & 2.85 \\
0.75 & 0.60 & 0.6 & 0.2 & 17.02 & 32.20 & 2.99 \\
0.75 & 0.60 & 0.8 & 0.0 & 13.87 & 30.43 & 2.99 \\
0.75 & 0.60 & 0.8 & 0.2 & 17.68 & 32.26 & 2.93 \\
0.75 & 0.60 & 1.0 & 0.0 & 14.29 & 31.33 & 2.70 \\
0.75 & 0.60 & 1.0 & 0.2 & 16.06 & 28.80 & 2.91 \\
\hline
0.90 & 0.50 & 0.0 & 0.0 & 7.45 & 22.19 & 1.816 \\
0.90 & 0.50 & 0.0 & 0.2 & 7.45 & 20.62 & 2.155 \\
0.90 & 0.50 & 0.2 & 0.0 & 12.01 & 26.76 &  2.49 \\
0.90 & 0.50 & 0.2 & 0.2 & 14.03 & 28.32 & 2.59 \\
0.90 & 0.50 & 0.4 & 0.0 & 13.43 & 30.97 & 2.94 \\
0.90 & 0.50 & 0.4 & 0.2 & 16.52 & 32.17 & 3.27 \\
0.90 & 0.50 & 0.6 & 0.0 & 13.47 & 27.77 & 3.86 \\
0.90 & 0.50 & 0.6 & 0.2 & 17.02 & 32.20 & 3.74 \\
0.90 & 0.50 & 0.8 & 0.0 & 13.87 & 30.43 & 3.73 \\
0.90 & 0.50 & 0.8 & 0.2 & 17.68 & 32.26 & 4.01 \\
0.90 & 0.50 & 1.0 & 0.0 & 14.29 & 31.33 & 4.24 \\
0.90 & 0.50 & 1.0 & 0.2 & 16.06 & 28.80 & 4.33 \\
\hline
\end{tabular}
}

\textbf{Notes.} $\rho$ is the ring particle internal density, $r_p$ is the ring particle radius, $\mu_r$ is the coefficient of static friction, $\mu_r$ is the coefficient of rolling friction, $\nu_{\mathrm{coll}}$ is the collisional viscosity (in units of $\Omega r^2_p$) , $\nu_{\mathrm{trans}}$ is the translational viscosity (in units of $\Omega r^2_p$), and $Q$ is the Toomre parameter.\\
\end{table}

\subsection{Correlation between Friction and Viscosity}
\begin{figure*}
\centering
\includegraphics[width=0.9\textwidth]{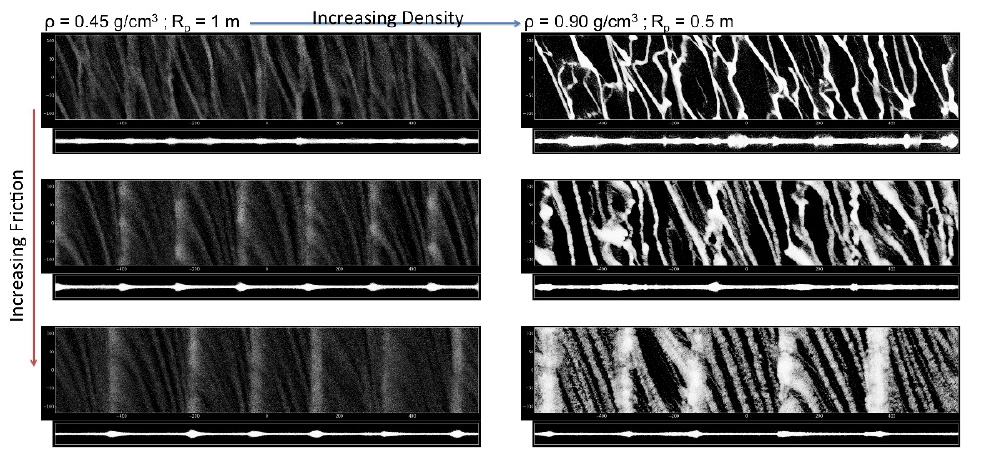} 
\caption{\footnotesize{Snapshots of some of our simulations after 100 orbits of collisional and gravitational evolution. The larger images show the position of ring particles in radial ($x$) and azimuthal ($y$) space. The smaller images show the positions of particles in radial and vertical ($z$) space. The left and right columns show ring particles with internal densities of 0.45 and 0.9 g/cm$^3$, respectively. When we allow the particles to interact with higher frictional forces (lower rows in the Figure have higher friction), we find that the ring particles are better at maintaining azimuthal ($y$-direction) axisymmetric features, brought about by the viscous overstability. For lower friction, the patch is not viscous enough to be stable against self-gravity wakes (inclined features most prominent in top panels).}}\label{ring_snapshots}
\end{figure*}

Figure~\ref{ring_snapshots} shows snapshots of our simulations after 100 orbits of collisional and gravitational evolution. We find that when particles have higher surface friction, axisymmetric features become more prevalent. Visual inspection of multiple frames allows us to ascertain that these features are likely brought about by viscous overstability wakes. At lower friction values, the ring particles form self-gravity wake structures that are inclined with respect to the orbital direction (inclined features most prominent in top panels of Fig.~\ref{ring_snapshots}).

For moderately high-friction cases, we see that the patch also reaches an equilibrium state with characteristic viscous overstability pulsations. Unlike previous simulations that studied the overstability (Salo et al.\ 2001), we see very little sign of gravitational wakes in this case even for a moderately high internal particle density of 0.45 g cm$^{-3}$. Salo et al.\ (2001) showed that the viscous overstability is present when ring particles have low internal density (0.20--0.30 g cm$^{-3}$). The axisymmetric features are less prominent at intermediate internal densities (0.45 g cm$^{-3}$), and likely non-existent for particles with densities of 0.90 g cm$^{-3}$. We find that, for sufficiently high surface friction, overstability features may exist even for non-porous water ice with densities of 0.90 g cm$^{-3}$ (see bottom-right panel of Fig.~\ref{ring_snapshots}).

\begin{figure}[h!]
\centering
\includegraphics[width=0.5\textwidth]{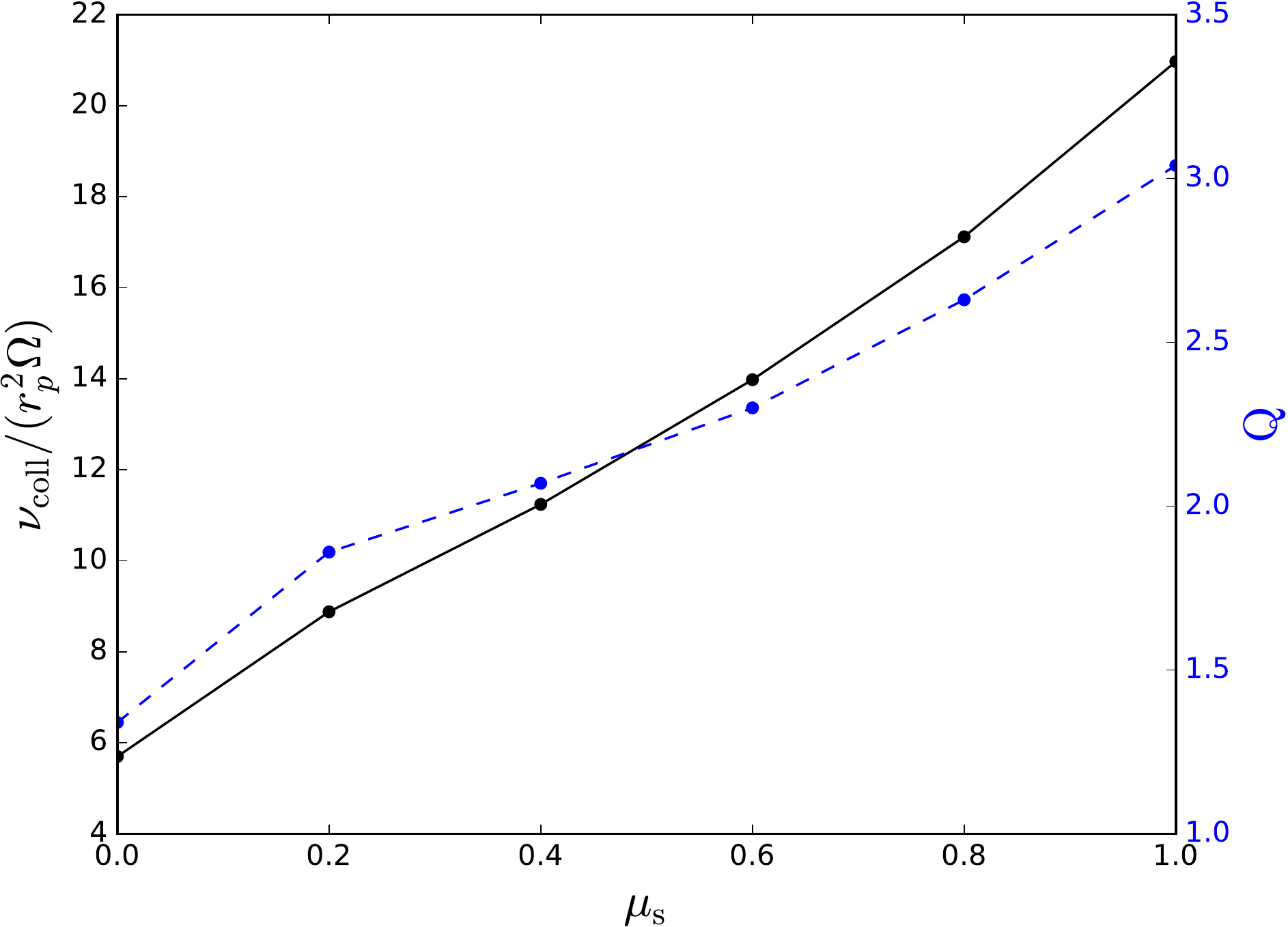} 
\caption{\footnotesize{Collisional viscosity of the patch (left axis) and Toomre $Q$ parameter (right axis) as a function of particle static friction $\mu_s$.  A high friction value increases the viscosity of the patch, and inhibits the formation of self-gravity wakes.}}\label{Fig3_Visc_Q}
\end{figure}

 Quantitatively, we can compare the change in the collisional viscosity of the patch with the Toomre $Q$ parameter, as a function of particle static friction $\mu_s$ (Fig.~\ref{Fig3_Visc_Q}). The $Q$ parameter is a ratio of the particle's radial velocity dispersion to the ring patch's surface density. Self-gravity wakes develop for $Q$ $<$ 2. A high friction value increases the viscosity of the patch and inhibits the formation of self-gravity wakes.

\subsection{Correlation with Particle Density}

We studied the dependence of the collisional viscosity with particle density. We find that for sufficiently large friction values (here $\mu_s$ = 1), the viscosity stabilizes the disk against self-gravity wakes even for very dense particles. Previous results that did not include explicit treatment of friction and multi-contact physics (Salo et al.\ 2001, Hedman et al.\ 2014) showed that the internal density of particles needed to be $<$ 0.30 g cm$^{-3}$ for viscous overstability to develop. Here we find that even particles with 0.90 g cm$^{-3}$ can exhibit the overstability.

\begin{figure}[h!]
\centering
\includegraphics[width=0.5\textwidth]{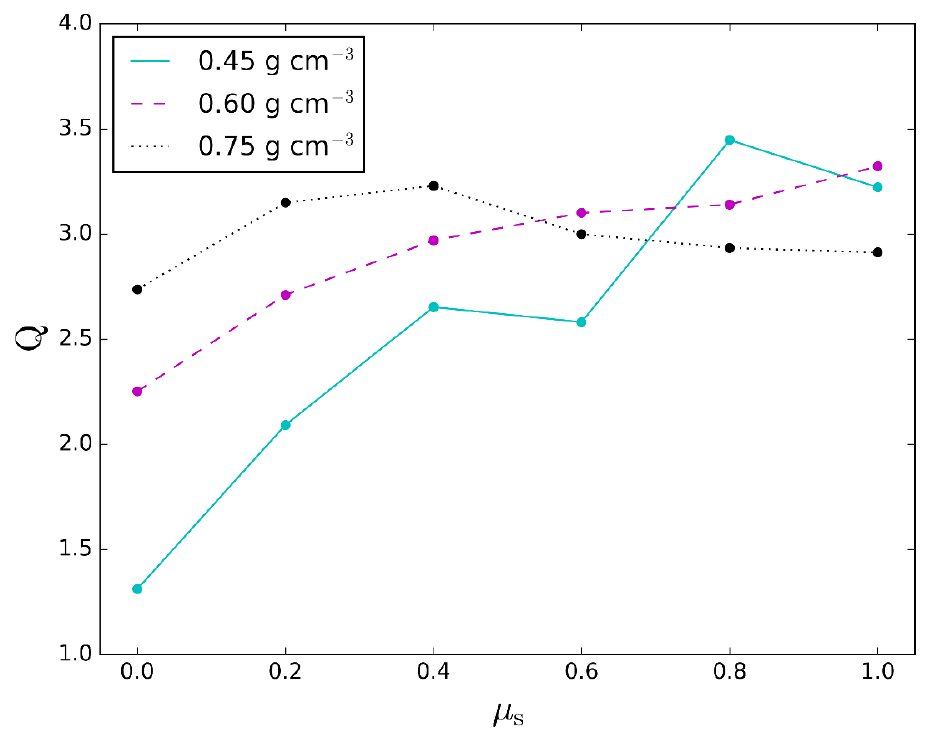} 
\caption{\footnotesize{The relationship between particle density (denoted by different linestyles and colors), the static friction component, and Toomre $Q$. These are cases where $\mu_{\mathrm{r}}$ = 0.2. The solid cyan line represents cases with $\rho_{\mathrm{p}}$ = 0.45 g cm$^{-3}$ and $R_{\mathrm{p}}$ = 1.0 m.  The dashed magenta line represents cases with $\rho_{\mathrm{p}}$ = 0.60 g cm$^{-3}$ and $R_{\mathrm{p}}$ = 0.75 m. The dotted black line represents cases with $\rho_{\mathrm{p}}$ = 0.75 g cm$^{-3}$ and $R_{\mathrm{p}}$ = 0.60 m. $Q$ increases with $\mu_s$ for almost all density cases, indicating a shift away from the dominance of self-gravity wakes.}}\label{Q_Density}
\end{figure}

 To further test the influence of particle density on large-scale structure formation, we plot the relationship between $\rho$, the friction parameters, and $Q$ in Fig.~\ref{Q_Density}. We show the value of $Q$ found in simulations with particle densities of 0.45, 0.60, and 0.75 g cm$^{-3}$, for values of $\mu_s$ ranging from 0.0--1.0, and for a rolling friction value of $\mu_r$ = 0.2. Here, we see that all particle densities exhibit a positive correlation between the particle friction values. The friction parameter here acts to increase the viscosity of the patch, and subsequently drives the formation of large-scale structure mainly generated by overstability wakes. At sufficiently high friction values, the value of $Q$ seems to saturate at $\sim$ 3. 
 
\subsection{Photometric Optical Depth Measurements}
Using the ray-tracing algorithm described in Section 2.3.2, we generated synthetic optical depth profiles of our library of simulations, in order to compare with Cassini observations.

By varying the azimuthal observing angle, $\theta$, we are able to investigate the nature of the large-scale structure in our simulations. Structures generated by the viscous overstability are axisymmetric; therefore, the photometric optical depth, $\tau_{\mathrm{phot}}$, is at a minimum when $\theta$ is aligned parallel to the azimuthal direction and at a maximum when $\theta$ is aligned parallel to the radial direction.

\begin{figure*}
\centering
\includegraphics[width=0.9\textwidth]{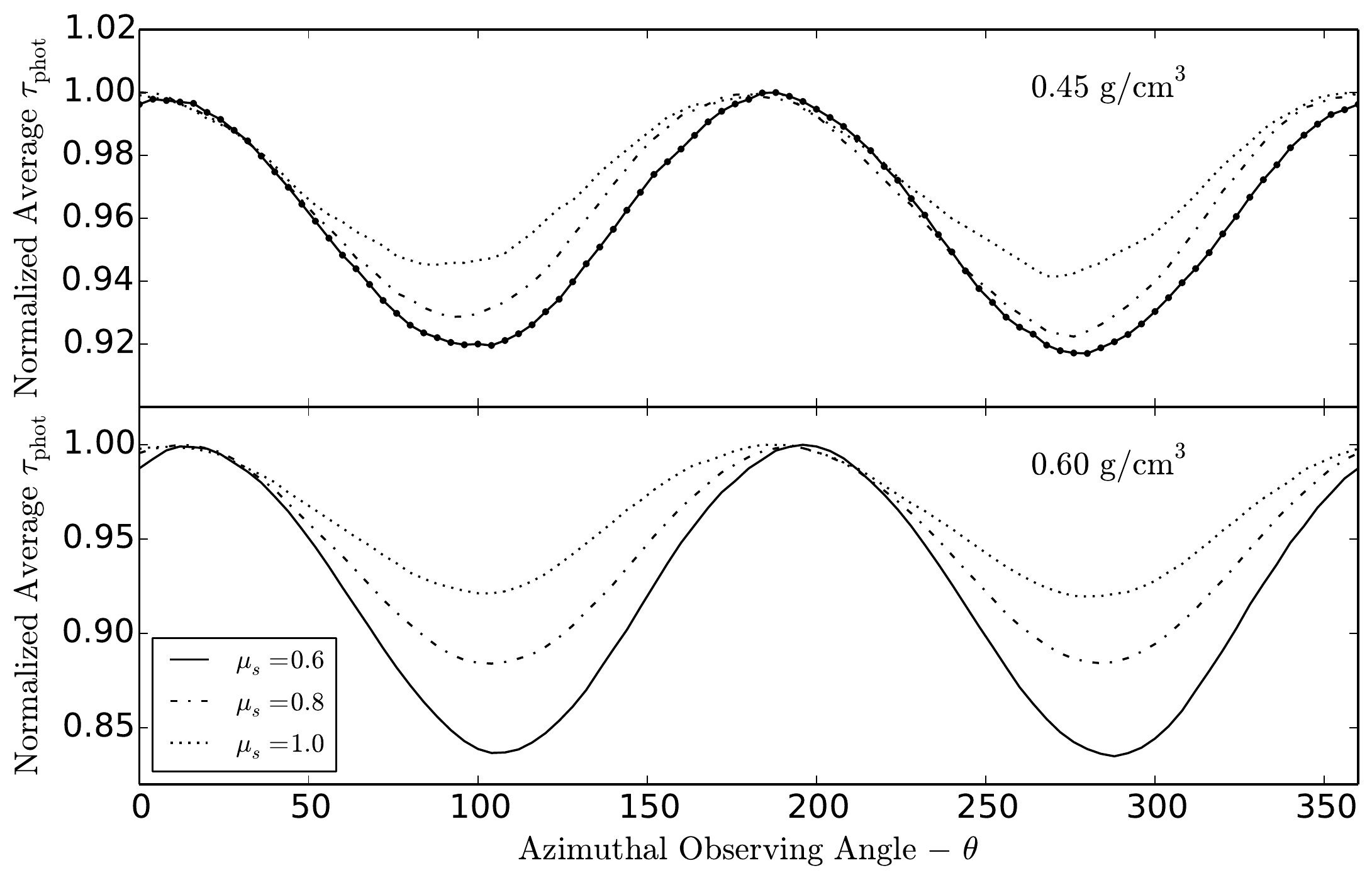} 
\caption{\footnotesize{An increase in the particle surface roughness evidently causes an observational shift from tilted self-gravity wake features to azimuthally coherent viscous overstability features. The average photometric optical depth is normalized by its maximum value, and it is plotted as a function of observing angle ($\degree$) at an elevation $B$ = 30$\degree$ for simulations with a particle density of 0.45 g cm$^{-3}$ (top panel) and 0.60 g cm$^{-3}$ (bottom panel). For both panels, the solid, dash-dotted, and dotted curves represent simulations where particles have static friction values of 0.6, 0.8, and 1.0, respectively. Individual data points are plotted on the solid line in the top panel to show the 4$\degree$ sampling frequency that was used for each curve. The unnormalized mean photometric optical depths for each case are described in the text.}}\label{TauTheta}
\end{figure*}

In Fig.~\ref{TauTheta}, we show our generated $\tau_{\mathrm{phot}}$ profiles for two different particle density cases, 0.45 and 0.60 g cm$^{-3}$. In these figures, $\theta$ = 0 corresponds to an observing angle aligned with the radial direction facing inwards towards Saturn. $\theta$ increases in a counterclockwise fashion from this initial geometry (looking down on the plane from the positive $z$ direction). For each simulation, $\tau_{\mathrm{phot}}$ is calculated for an elevation angle of $B=30\degree$, and for $\theta$ ranging from 0 to 360$\degree$ sampled at 4$\degree$ intervals. In the presence of wake-like features, intermediate values of $B$ result in variations of $\tau_{\mathrm{phot}}$ with $\theta$. As $B$ approaches $90\degree$, the viewing geometry is less influenced by changes in $\theta$, resulting in a near-constant  $\tau_{\mathrm{phot}}$. In increasing static friction coefficient order (0.6 to 1.0), the mean photometric optical depths have values of 0.86, 0.95, and 1.03 for the 0.45  g cm$^{-3}$ case and 0.45, 0.64, and 0.73 for the 0.60  g cm$^{-3}$ case. For both cases, the mean photometric optical depth increases with the friction coefficient. Furthermore, for both particle density cases, we see that the amplitude of the $\tau_{\mathrm{phot}}$ variation decreases as a function of the friction parameters. These trends of the optical depth suggest that friction acts against self-gravitational clumping. While clumping causes the ring patch to become less opaque, an increase in particle friction allows the ring particles to diffuse away from high density regions, making the opacity larger and less variable. For most of the cases seen here for both particle densities, we see that the minima are located at $\theta$ values between $\sim$ 90$\degree$ and 110$\degree$. This is consistent with values of $\sim$ 90$\degree$ to 120$\degree$ for km-scale structure found in Saturn's dense rings (e.g., Colwell et al.\ 2007). The observed value is 90$\degree$ for the outer B ring, indicating existence of overstability features. Although the pitch angles seen in all of our simulations are consistent with observations, the average values of $\tau_\mathrm{phot}$ we measure ($\sim$ 1) are much lower than those for the outer B ring, which are a few factors larger than 1 and can be greater than 5. It is possible that inter-wake gaps are filled by small particles that are not included in our simulations, but that would contribute much of the observed opacity. 

We also find that for higher friction values, the minima are shifted towards 90$\degree$, or axisymmetry. This is true for both density cases. In order to obtain an accurate estimate of the azimuthal angle that coincides with a given minimum $\tau_{\mathrm{phot}}$, a sinusoidal curve that best fits the data was generated using a least-squares minimization technique.  For the case of 0.45 g cm$^{-3}$, we find minima at values of 100.1$\degree$, 93.6$\degree$, and 88.5$\degree$ for $\mu_s$ = 0.6, 0.8, and 1.0, respectively. For the case of 0.60 g cm$^{-3}$, we find minima at values of 109.5$\degree$, 103.8$\degree$, and 95.8$\degree$ for $\mu_s$ = 0.6, 0.8, and 1.0, respectively. The shift is more apparent in the top panel of Fig.~\ref{TauTheta}, the lower-density case. This is expected, as larger particle densities tend to lead to stronger self-gravitational wakes. Hence, for similar friction values, lower-density particles are more strongly driven to axisymmetry and viscous overstability.

\subsection{Comparison with Cassini UVIS occultations}
\begin{figure}[h!]
\centering
\includegraphics[width=\columnwidth]{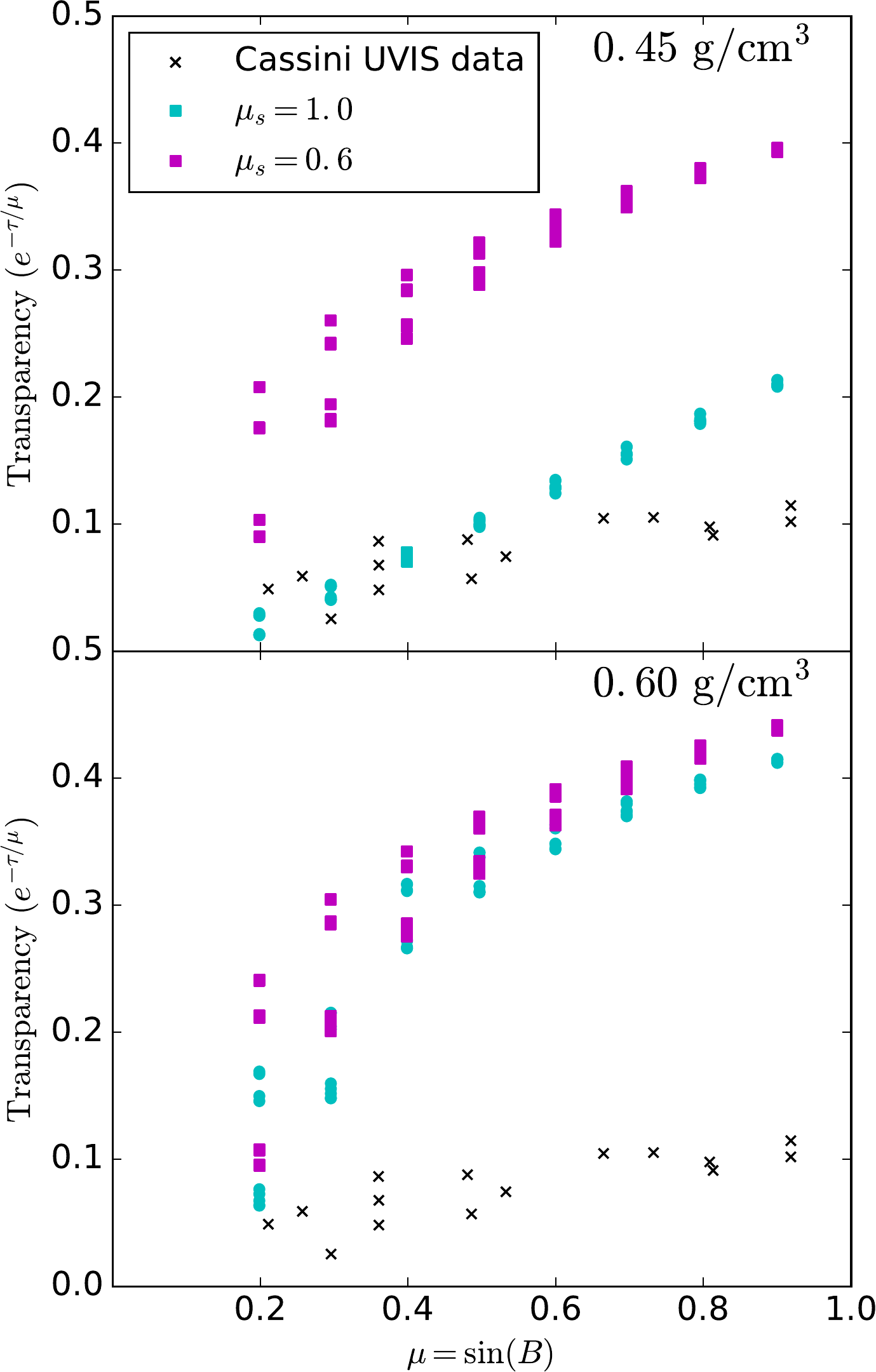}
 \caption{\footnotesize{The transparency of the simulated data (circles and squares) is compared to Cassini UVIS stellar occultation data (crosses). The top and bottom panels plot transparency of two friction cases (cyan circles for $\mu_s$ = 1.0 and magenta squares for $\mu_s$ = 0.6) as a function of the projection factor for simulations with particle densities of 0.45 g/cm$^3$ (top panel) and 0.60 g/cm$^3$ (bottom panel), respectively. We find that higher values of friction lead to lower ring patch transparency.}}\label{TransparencyMu}
 \end{figure}

Finally, we compare the results of our simulations to Cassini UVIS stellar occultations (Colwell et al.\ 2006). Observations of stellar occultations by the UVIS instrument reveal the transparency of the ring as a function of observation geometry. These observations vary in both the azimuthal observing angle, $\theta$, and the spacecraft elevation angle $B$. Therefore, in order to effectively compare our simulation data to observations, we calculate the average transparency of our simulated patch for unique combinations of $\theta$ and $B$. The transparency, $T$, of the ring patch is defined as
\begin{equation}  
T = e^{-\tau_\mathrm{phot}/\mu},
\end{equation}
where $\mu$ is the projection factor, $\mu=|\sin(B)|$. In Fig.~\ref{TransparencyMu}, we plot $T$ as a function of $\mu$ for two particle densities and two different values of static friction $\mu_s$. For each value of $\mu$, we plot results for a range of the azimuthal observing angles, $\theta$. As expected, for larger elevation angles, $B$, the range of transparencies becomes smaller across the range of $\theta$. Overall, we find that a simulation where particles have lower densities and higher friction ($\rho$ = 0.45 g/cm$^3$ and $\mu_s$ = 1.0) is the best fit to the observations. Furthermore, in general, a higher value of static friction results in lower transparencies. 

%
%
\section{Conclusions \& Future Work}
We presented ring simulations that reveal the relationship between particle friction and large-scale structure in Saturn's B ring. A comprehensive explanation of the interplay between the mechanisms of self-gravity and viscous overstability has remained elusive. It is understood that both play an important role in generating large-scale structure in Saturn's rings. Here we have shown that it is possible that the surface properties of particles, here modeled through two friction parameters (static and rolling friction), may control whether a patch of ring particles exhibits self-gravitational wakes or axisymmetric overstable features, or both.

We find that higher interparticle friction increases the viscosity (collisional and translational), such that the ring particles become stable against self-gravitational perturbations. Furthermore, we have shown that the presence of axisymmetric features via viscous overstability does not necessarily require highly porous particles. In fact, non-porous particles may generate viscous overstability features if their surface roughness is sufficiently high.

This has important implications for the possible mass of the B ring itself. Previous studies (Robbins et al.\ 2010) have relied on similar simulations to estimate the masses of the A and B rings. Robbins et al.\ (2010) suggested an upper limit for the ring particle density of 0.45 g/cm$^3$, based on comparisons of simulation data to ring opacity measurements. Our simulations predict a similar ring particle density; however, they did not include the effects of inter-particle friction. We show that if highly frictional particles exist in the B ring, then B ring regions that exhibit axisymmetric features may be produced by particles with densities as high as 0.90 g/cm$^3$. Taking this into account, the current values for the mass of the B ring may be underestimated. 

We have performed comparisons to Cassini UVIS stellar occultation data and found that particles with lower densities and higher friction produce ring transparencies that best match the observations. This suggests that axisymmetric features are likely prevalent in the B ring, as both conditions of low densities and high friction are conducive to the generation of viscous overstability. However, the average photometric optical depths measured for these simulations ($\sim$ 1) are much lower than those observed in the B ring ($>$ 4 in some locations). If we were to simulate a ring patch with a size distribution of particles (and with more small particles in particular), the transparencies would likely be lower for all our cases. There have been indications that having a particle size distribution leads to more opaque rings compared to a monodisperse size distribution for rings made of particles with densities of 0.45 g/cm$^3$ (Robbins et al.\ 2010). However, this effect was not as pronounced as that in simulations with lower particle densities. Nevertheless, this suggests a possibility that particles with densities greater than 0.45 g/cm$^3$ may still be able to produce features that are sufficiently opaque to match observations, with the caveat that the particles would need to have high surface roughness.

Since friction plays an important role in controlling the macroscopic properties of the ring, our simulations have shown that understanding the surface roughness of the ring particles is essential if observations of large-scale structure are to be used to interpret their physical properties (size, densities, etc.). Here, we have chosen to explore a wide range of the friction parameter space in order to understand the range of possible behaviors. The actual surface roughness of ice particles in a vacuum are unknown. So far, simulations have relied on experimental work that were performed on non-rotating ice spheres that were either perfectly smooth (Bridges et al.\ 1984) or had a compacted frosty outer layer (Hatzes et al.\ 1988). There have been some experiments that have measured the friction properties of ice on ice. Sukhorukov \& L{\o}set (2013) conducted field tests on the friction of sea ice on sea ice in the Barents sea in temperatures of 253 to 271 K, and found that the coefficient of static friction can range from $\sim$ 0.6 to 1.26 depending on the length of time the ice is held together before sliding is allowed to begin (higher hold times result in higher friction). Schulson \& Fortt (2012) measured the coefficient of sliding friction of freshwater ice sliding slowly (with speeds up to 1 mm/s) at temperatures of 98 to 263 K. They found that the sliding friction can range from 0.15 to 0.76 depending on the temperature and sliding speed. While these results can provide some ground truth to numerical studies, the environmental conditions they were conducted in do not match that found around Saturn (vacuum at temperatures that range from 50 to 100 K for the B ring, Flasar et al.\ 2005) . Therefore, future experimental and theoretical studies should explore the collisional interaction of realistic ice particles in order to place better constraints on their surface properties.

Future work should also consider a size distribution of particles (within a single simulation) and a wider range of initial geometric optical depths in order to better quantify the effect that particle surface roughness has on observable properties of the rings. Furthermore, recent studies (Hedman et al.\ 2014) have shown that axisymmetric features in the A ring are azimuthally coherent on scales longer than those simulated here. If inter-particle friction does indeed allow high-density particles to form overstabilities, then simulation with much longer azimuthal extents will have to be performed to show that such developed structures are stable in the long term (1,000's of orbits). 

Finally, we have been studying the effect of inter-particle cohesive forces on the formation of km-scale structure. Tremaine (2003) showed that cohesion can have an important role in creating structure in Saturn's rings and predicted the formation of shear-free ringlets. We are exploring the extent to which cohesive forces can enhance or diminish the generation of self-gravitational wakes and axisymmetric features, or if they generate separate unique large-scale features. 

\section{Acknowledgements}
This material is based on work supported by the U.S. National Aeronautics and Space Administration under Grant no.\ NNX14AO36G. The authors would like to thank the anonymous referee for their valuable feedback. Simulations were performed on the YORP cluster administered by the Center for Theory and Computation, part of the Department of Astronomy at the University of Maryland; the Deepthought2 HPC cluster at the University of Maryland; and the MARCC Cluster run jointly by Johns Hopkins University and the University of Maryland.


\end{document}